\documentclass[english,preprint,preprintnumbers,amsmath,amssymb,superscriptaddress]{revtex4}
\usepackage{verbatim}
\usepackage{graphicx}
\usepackage{amssymb}
\usepackage{babel}
\usepackage{color}
\makeatother

\newcommand{\bra}[1]{\left\langle {#1} \right|}
\newcommand{\ket}[1]{\left|  {#1} \right\rangle}

\newcommand{\abs}[1]{\left| {#1} \right|}

\begin{document}
\title{A controllable single photon beam-splitter as a node of a quantum network}
\author{Gaurav Gautam}
\thanks{These two authors contributed equally}
\affiliation{Department of Physics, Indian Institute of Technology, Kanpur 208016, India}
\author{Santosh Kumar}
\thanks{These two authors contributed equally}
\affiliation{School of Physical Sciences, Jawaharlal Nehru University, New Delhi 110067, India}
\affiliation{Homer L. Dodge Department of Physics and Astronomy, The University of Oklahoma, 440 W. Brooks St. Norman, OK 73019, USA}

\author{Saikat Ghosh}
\email{gsaikat@iitk.ac.in}
\affiliation{Department of Physics, Indian Institute of Technology, Kanpur 208016, India}
\author{Deepak Kumar}
\email{dk0700@mail.jnu.ac.in}
\affiliation{School of Physical Sciences, Jawaharlal Nehru University, New Delhi 110067, India}

\date{\today}

\begin{abstract}
A model for a controlled single-photon beam-splitter is proposed and analysed. It consists of two crossed optical-cavities with overlapping waists, dynamically coupled to a single flying atom. The system is shown to route a single photon with near-unity efficiency in an effective ``weak-coupling'' regime. Furthermore, two such nodes, forming a segment of a quantum network, are shown to perform several controlled quantum operations. All one-qubit operations involve a transfer of a photon from one cavity  to another in a single node, while two-qubit operations involve transfer from one node to a next one, coupled via an optical fiber. Novel timing protocols for classical optical fields are found to simplify possible experimental realizations along with achievable effective parameter regime. Though our analysis here is restricted to a cavity-QED scenario, basic features of the model can be extended to various other physical systems including gated quantum dots, circuit-QED or opto-mechanical elements.

\end{abstract}

\pacs{ 32.80.-t, 42.50.-p, 42.81.-i, 42.81.Qb}

\maketitle

\section{Introduction}
Exploring the potential of quantum systems as a novel and powerful resource for information processing and communication has  been a major focus of research over past three decades\cite{NielsenChuang,BEZ,Everett,Zoller05,Ladd10}.  In general, quantum information operations are usually envisioned on a network of nodes \cite{Cirac99,DiV2000,BA04,Kimble08} consisting of material media while the nodes in turn exchange information via a network of photonic bus consisting of single photons. A variety of physical systems have been explored as possible candidates towards realizing these effective nodes which includes ion-chains \cite{BW08,DM10}, optical  qubits \cite{KLM01,BF09,RP10,AHT13}, Josephson junction qubits \cite{DWM04,Sidd11}, spin and charge qubits in quantum dots \cite{LDiV98,Petta05} and electromagnetic cavities with trapped atoms \cite{CZKM97,DK04,Ritter12}. For a quantum network, while single photons connect these nodes robustly, the nodes perform physical operations including storing \cite{Haruka09} and routing of photons \cite{Aoki09,Hoi11,Ma11,AH12,Zhou13}, entangling remote locations \cite{DLCZ,Kimble08} or performing basic gate operations \cite{NielsenChuang,BEZ}. 

In particular, there are several recent proposals and experimental realizations of controlled photon routers. Some of these include experimental demonstration of micro-fabricated optical cavities with single atoms efficiently coupling input-output modes of a one-dimensional tapered fiber \cite{Aoki09} or redirecting single and bi-photons in different output modes \cite{Tiecke14}, superconducting transmon qubits redirecting microwave photons in separate outputs using quantum interference \cite{Hoi11} or proposals of opto-mechanical systems routing photons forward or backward \cite{AH12} or coupled resonator waveguides routing photons in orthogonal directions \cite{Zhou13}. 

Here we propose a model that can act as a versatile quantum node in a network. When operated as a beam splitter, the splitting amplitudes can be fully controlled along with routing a single photon in orthogonal directions with near unity quantum efficiency. Furthermore, two such nodes can be combined for gate operations with one and two qubits. 

The system consists of a basic unit of two crossed optical cavities with overlapping cavity waists along with an atom at the waist. We show that it can transfer coherently a single photon from one cavity mode to the other via an adiabatic rotation of a dark state triggered by either a passage of a single atom through the cavity waist along with timed classical optical pulses or by using sequence of laser pulses along with a static atom at the waist.

The proposed scheme works for small values of the cavity coupling constant $g$ compared to classical fields and is effective in the weak-coupling limit. For the \textit{photon-router regime}, there is parameter range for which one can get almost unity routing efficiency with minimal losses. For the more general \textit{beam-splitter regime}, we derive an effective Hamiltonian and the corresponding time independent unitary 
connecting the two initial photonic(or cavity) modes to the transformed ones, by adiabatically eliminating the atom, weakly coupled to the cavities. The resulting analytic expression for the unitary agrees well with an exact simulation of the full system. 

This crossed-cavity system can also serve as a node of a future quantum network. For a choice of two internal states of the atom as a logical qubit, photon-routing corresponds to unitary operations on the qubit. The unit can then be repeated with optical fibers as interconnects to construct a network. An analysis for two such nodes is presented here. It is found that the presence of a photon in a node allows arbitrary unitary operations on the corresponding local qubit without affecting the other node in certain parameter regimes. Further there are operations through which a photon can be adiabatically transferred from one node to the other which enables certain two-qubit operations such as generation of entangled atomic states. Thus via inter-node transfer of photon in conjunction with single node processes on the node in which the photon is present, arbitrary one-qubit and two qubit operations can be carried out in a controlled manner. All such operations on qubit processing can then be implemented with protocols on classical field timing and power. We present examples of such protocols for some representative operations. Finally, effect of losses is analysed and it is found that there is a suppression of loss in all the processes considered here.

\section{Physical System}
The proposed model has the following physical elements: (a) two single mode optical cavities (labelled $l$ and $r$ hereafter) geometrically arranged such that the cavity axes are orthogonal to each other while their waists overlap (see Fig.1). (b) A single atom, prepared in a specific state of its ground-state manifold at the overlapping modes of the cavities. We consider two distinct methods of atom-cavity coupling: the coupling is either dynamically induced as the atom is physically transported (either with an optical tweezer, an optical \textit{conveyor belt} or simply, the atom flying across at a constant velocity) through the overlapping waist or a stationary atom trapped at the cavity waist. While in the first case, the atom-cavity coupling constant $g$ depends explicitly on time, in the latter it is a constant. (c) Two classical fields or lasers with specific polarizations and frequencies, resonant with the atom.

The corresponding Hamiltonian for independent cavities and the atom (considering a single resonant mode for each cavity) is of the form:
\[
\hat{H}_0 =  \sum_{i=l,r} \hbar \omega_i \left( \hat{C}_i^{\dagger} \hat{C}_i + \frac{1}{2} \right) 
+ \displaystyle\sum_{x} \hbar \epsilon_x \vert x \rangle \langle x \vert
\] 
where $\hat{C}_i^{\dagger}, i=l,r$ are creation operators for cavity modes with frequencies $\omega_i$, and  $\ket{x}$ denotes the atomic levels with energy $\hbar \epsilon_x$ (which also includes the shifts due to far off-resonant dipole trapping fields for transporting the atom).

At the overlapping waist the atom interacts with the fields of each of the cavities, absorbing or emitting single cavity photons. Each of the cavity modes are resonant to an energy corresponding to an atomic transition between the ground state  $\ket{f_i}$ and excited state $\ket{e_i}, i=l, r$. Furthermore, two classical fields couple the excited states $\ket{e_l}$ and $\ket{e_r}$ to a single ground state $\ket{f_m}$ (see Fig.1b). The atom-field interaction can then be effectively described with only five of these atomic levels \cite{Pellizzari97,Simon07}. Within the \textit{dipole approximation}, the effective interaction Hamiltonian then takes the form:
\[
\hat{H}_I = - \sum_{i=l,r} \hbar \left[ g_{i} \hat{C}_{i} \ket{e_i}\bra{f_i} + \Omega_i e^{-i \nu_{i} t} \ket{e_i}\bra{f_m} \right] + h.c. 
\] 

The first term on the right describes interaction of the atom with modes of the cavities with coupling constant $g_i$'s while the last term describes the classical laser fields of \textit{Rabi} frequencies $\Omega_i$ interacting with the atomic levels $(\ket{f_m},\ket{e_i})$. 

With the resulting geometry, the quantization axis should be uniquely chosen for describing states of the atom along with polarizations of the classical and cavity fields. In the present geometry, since the two cavities are orthogonal, there might be ambiguity in interpreting the polarizations and realizing the interaction Hamiltonian. Next, we argue that the above interaction Hamiltonian is indeed realizable with a judicious choice of the quantization axis and classical field directions.
 
\begin{figure*}
\includegraphics[scale=0.7]{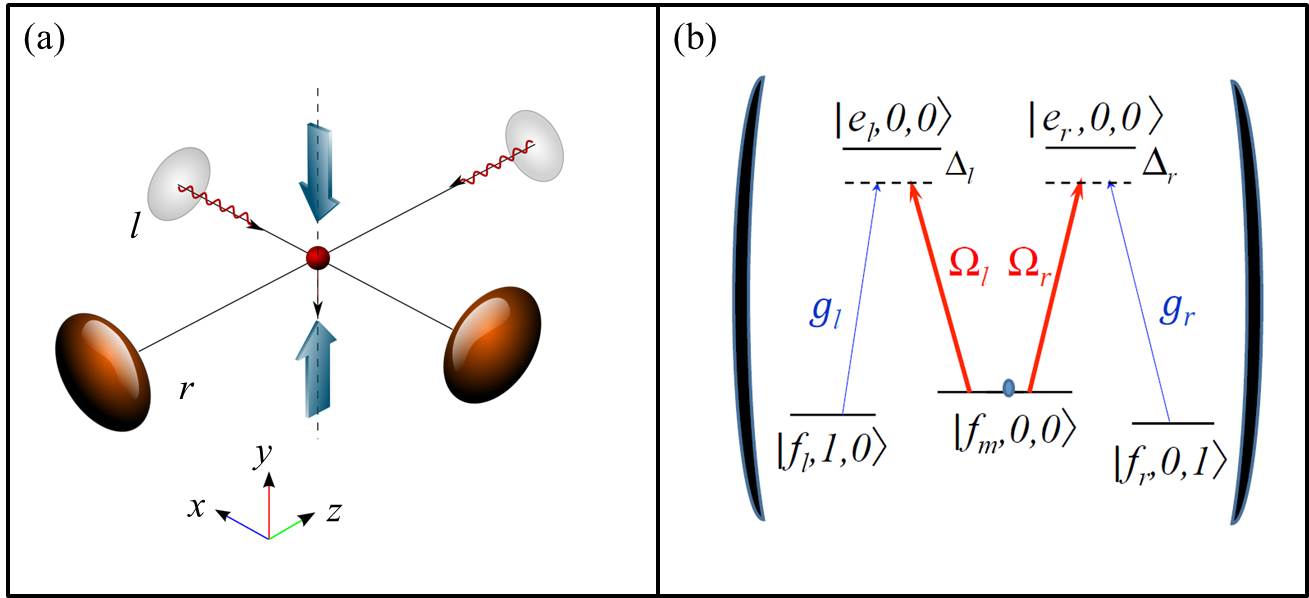}
\caption{(a) A cartoon depicting the physical model, with two optical cavities (labelled $l$ and $r$) and an atom at the overlapping waist of the two cavities. Along with classical optical fields (thick arrows orthogonal to the cavity plane), the model acts as a controllable beam splitter for a single photon. (b) In the single excitation regime, the model is effectively described with five states. State $\ket{f_l, 1, 0}$ ($\ket{f_r, 0, 1}$) is a state with a single photon in cavity $l$ ($r$) while the atom is in a ground state $\ket{f_l}$($\ket{f_r}$). States $\ket{e_l}$, $\ket{e_r}$ are atomic excited states which are coupled to a common ground state $\ket{f_m}$ and none of these three states contain any photon in the cavity. $g_l$ ($g_r$) is the $l$($r$) cavity coupling. Cavity modes are two-photon resonant with the classical field $\Omega_l$ ($\Omega_r$) and detuned from the excited state by $\Delta_l$ ($\Delta_r$).}
\end{figure*} 

\textit{Quantization axis:}  For a choice of the quantization axis  along the axis of cavity $l$ ( $z-axis$, Fig.1(a)) and the two classical fields $\Omega_l,\Omega_r$ along the $y-direction$ i.e. orthogonal to the plane of the two cavities, the atom can be initialized to a state $\ket{f_l} = \ket{F=1, m_F = -1}$ (corresponding to a hyperfine ground state of an atom) such that cavity $l$ is resonant (or near resonant with a detuning $\Delta_l$) to a hyperfine excited state $\ket{F'=1}$ (atomic state symbols having usual meaning). A right-circularly polarized $\hat{\sigma}_+$ photon in the cavity will then induce a transition to the excited state $\ket{e_l} = \ket{F'=1, m_F' = 0}$. To couple the levels $\ket{f_m} = \ket{F=1, m_F = 0}$ and $\ket{e_l}$, the classical field $\Omega_l$ needs to be $\hat{\pi}$ polarized i.e. polarized along $z-axis$, while propagating along the $y-axis$. 
Since the cavity $r$ is tuned between the levels $\ket{f_r} = \ket{F=1, m_F' = 1}$ and $\ket{e_r} = \ket{F'=2, m_F' = 0}$, the cavity photon polarization corresponds to $\hat{\sigma}_{-}$ or equivalently, a linearly polarized field along the $y$ direction. Finally, the second classical field $\Omega_r$, resonant between $\ket{f_m}$ and $\ket{e_r}$, needs to be polarized along the $x$-axis, which is the axis of cavity $r$. Accordingly, these choices for polarizations and states can realize the total Hamiltonian.

\section{Analysis} The model is effectively described with a ``dressed'' state
\[
\ket{\psi(t)} = \sum_{x,n_l,n_r} e^{-i(\epsilon_x + n_l \omega_l + n_r \omega_r)t} A_{x}(n_l,n_r,t) \ket{x, n_1, n_2}. \label{OpticalBlochEq}
\]
Here the state $\ket{x,n_l,n_r}$ is in the joint space of the atom along with $n_l$ and $n_r$ photons in the two cavities with an amplitude $A_{x}(n_l,n_r,t)$. The conservation of the excitation number M ( $=$ number of atomic excitations in $e_l, e_r, f_m$ + total photon number) allows one to work in a specific sector with fixed number of excitations, decoupling other such sectors completely (ignoring decay and losses). 

Equations for the probability amplitudes are then,
\begin{eqnarray}
-i \dot{A_{f_l}}(n_l +1,n_r,t) &=& g_l^* \sqrt{n_l +1} A_{e_l}(n_l, n_r,t)  e^{-i(\Delta_{l} + \Delta_{g_l})t}, \\
-i \dot{A_{f_r}}(n_l,n_r +1,t) &=& g_r^*  \sqrt{n_r +1} A_{e_r}(n_l,n_r,t) e^{-i(\Delta_{r} + \Delta_{g_r})t}, \\
-i \dot{A_{f_m}}(n_l,n_r,t) &=& \Omega_l^{*} A_{e_l}(n_l,n_r,t) e^{-i\Delta_{l}t} + \Omega_r^{*} A_{e_r}(n_l,n_r,t) e^{-i\Delta_{r}t}, \\
-i \dot{A}_{e_l}(n_l,n_r,t) &=& g_{r} \sqrt{n_l +1} A_{f_l}(n_l + 1, n_r,t) e^{i(\Delta_{l} + \Delta_{g_l})t} + \Omega_l A_{f_m}(n_l,n_r,t) e^{i\Delta_{l}t}, \\
-i \dot{A}_{e_r}(n_l,n_r,t) &=& g_2 \sqrt{n_r +1} A_{f_r}(n_l ,n_r+1,t) e^{i(\Delta_{r} + \Delta_{g_r})t} + \Omega_r A_{f_m}(n_l,n_r,t) e^{i\Delta_{r}t},
\end{eqnarray}
where $\epsilon_{e_l}-\epsilon_{f_l}-\omega_l = \Delta_{l} + \Delta_{g_l}, \epsilon_{e_r}-\epsilon_{r}-\omega_r = \Delta_{r} + \Delta_{g_r}, \epsilon_{e_l}-\nu_{}-\epsilon_{f_m} = \Delta_{l}, \epsilon_{e_r}-\nu_{r}-\epsilon_{f_m} = \Delta_{r}.$  with $\Delta_{l}$ and $\Delta_{r}$ as the detuning parameters. Using standard transformation,
$\tilde{A}_{e_i} = A_{e_i} e^{-i \Delta_{i} t}, \tilde{A}_{i} = A_{i} e^{-i \Delta_{g_i} t}$ for  $i = l, r$ and $\tilde{A_{g}} = A_{g}$, equations assume the form, 
\begin{eqnarray}
-i \frac{\partial \tilde{A}(t)}{\partial t} = M \tilde{A}, 
\end{eqnarray} where
\begin{eqnarray} 
 M = \left( \begin{array}{ccccc}
              \Delta_{g_l} &  0 & 0 & g_{l_n}^* & 0\\
		0  &   \Delta_{g_r} & 0 & 0 & g_{r_n}^*\\
	0 & 0 &  0 & \Omega_{l}^{*} & \Omega_{r}^{*} \\
	 g_{l_n} & 0 & \Omega_{l} & -\Delta_{l} & 0 \\
	  0 &  g_{r_n} & \Omega_{r} & 0 & -\Delta_{r} \\
	  \end{array} \right),
\end{eqnarray}
$g_{l_n} =\sqrt{n_1 +1} g_{l}$ and $g_{r_n} =\sqrt{n_r +1} g_{r}$. 
Under the condition of Raman resonance, $\Delta_{g_l}=\Delta_{g_r}=0$, the system has a \textit{dark state}
\begin{equation}
\ket{D_{(n_l,n_r)}} = \ket{f_m,n_l, n_r} - \frac{\Omega_l}{g_{l_n}}\ket{f_1,n_l+1,n_r} - \frac{\Omega_r}{g_{r_n}}\ket{f_2,n_l,n_r+1}.
\end{equation}
free of any of the excited states $\ket{e_i}$. An atom in such a state therefore does not emit any photons or remains \textit{dark}. 

This dark state $\ket{D_{(n_l,n_r)}}$ captures some of the essential physics of our model. In particular for the single-photon sector (corresponding to $n_l = n_r = 0$), the state $\ket{D_{(0,0)}}$ can be $adiabatically$ rotated from an initial state $\ket{f_l,1,0}$ to a final state $\ket{f_r,0,1}$ by ramping up (down) and down (up) the fields $\Omega_l$ ($\Omega_r$), as long as one is slower than any other relevant time-scales in the system.  Most importantly, one can note that for this model, the initial and the final states correspond to a photon in each of the cavities or equivalently, the adiabatic rotation can route a single photon in an orthogonal direction. 

Similar dark states and corresponding adiabatic rotation with \textit{counter-intuitive} pulse sequences has been well studied and applied in varying physical scenarios. However, one can note that for the present model, precise timing requirement of such dark state rotation with counter-intuitive pulse sequences can get too demanding. 

On the contrary, as we show next, there is a particularly interesting and yet unexplored off-resonant regime corresponding to $\Delta_i \gg \Omega_i \gg g_i$. In this regime, the photon can not only be routed from one orthogonal mode to the other with unity efficiency but can also be put into arbitrary superposition of two cavity modes by changing a single parameter in the model and with a simple, intuitive sequence of pulses. 

Since the physics in all $M$-sectors is similar, for simplicity, we present an analysis of $M=1$ i.e. sector corresponding to single excitation. 

\textit{Large-detuning limit: Adiabatic rotation}
In the large detuning limit ($\Delta_{l}=\Delta_{r}=\Delta \gg \Omega_i ,g_i$ ) excited states are effectively decoupled from the dynamics (Eqs. (4) and (5)), and can be adiabatically eliminated.  The system is then well described with an effective 3-level Hamiltonian:

\begin{equation} \label{3levelH}
H_{eff}=
\begin{bmatrix}
\frac{\abs{\Omega_{l}}^{2}}{\Delta_l}+\frac{\abs{\Omega_{r}}^{2}}{\Delta_r} & \frac{\Omega_{l}^{\ast}g_{l}}{\Delta_l} & \frac{\Omega_{r}^{\ast}g_{r}}{\Delta_r} \\
\frac{\Omega_{l}g_{l}^{\ast}}{\Delta_l} & \frac{\abs{g_{l}}^{2}}{\Delta_l} & 0 \\
\frac{\Omega_{r}g_{r}^{\ast}}{\Delta_r} & 0 & \frac{\abs{g_{r}}^{2}}{\Delta_r} 
\end{bmatrix},
\end{equation}
in the basis states $\left[ \ket{f_m,0,0} = \ket{F_m}, \ket{f_l,1,0} = \ket{F_l}, \ket{f_r,0,1} = \ket{F_r} \right]$. While all the states pick up light-shifts due to their coupling to the excited states (diagonal terms), they also get coupled to each other through an effective two-photon Rabi frequency.
$s_l=\frac{\Omega_{l}^{\ast}g_{l}}{\Delta_l}$ ($s_r=\frac{\Omega_{r}^{\ast}g_{r}}{\Delta_r}$) couples the state with one photon in the cavity, $\ket{F_l}$ ($\ket{F_r}$) to the state $\ket{F_m}$ with the atom transiting to the state by absorbing the cavity photon.

The Hamiltonian $H_{eff}$ again has a dark state consisting of only the dressed states $\ket{F_l}=\ket{f_l,1,0}, \ket{F_r}=\ket{f_r,0,1}$, given by
\[
 \ket{D_e} = s_r \ket{F_l} - s_l \ket{F_r}
\].

It can be heuristically argued that the evolution of the intermediate ground state, to first order, is driven by the two excited states (we verify this argument in the next section with numerical simulations). Accordingly, for $\Omega_i << \Delta$ one can use $\dot{A}_{f_m} \simeq 0$ to eliminate  the intermediate state. This gives rise to an effective Hamiltonian (keeping the two detunings $\Delta_l$ and $\Delta_r$ explicitly)of the form:

\begin{equation} \label{BS}
H_{BS}=
\begin{bmatrix}
-\frac{\abs{g_l}^2}{\Delta_l} +\frac{\abs{s_l}^2}{\delta_m} & \frac{s_l^{\ast}s_r}{\delta_m} \\
\frac{s_{l}s_{r}^{\ast}}{\delta_m} &
-\frac{\abs{g_{r}}^2}{\Delta_r} +\frac{\abs{s_{r}}^{2}}{\delta_m},
\end{bmatrix}
\end{equation}

where

\begin{equation}\label{delta-m}
\delta_m = \sum_j\frac{\abs{\Omega_j^2}}{\Delta_j}.
\end{equation}

It can be noted that this Hamiltonian is akin to a beam-splitter like interaction, mixing the two orthogonal optical modes. 

\textit{Numerical Results:}

To find the effective parameter regime for the validity of the heuristic argument, the effective five-level Hamiltonian has been numerically simulated. The results are summarized in Fig. 2 and Fig. 3. For these and all subsequent numerical results, the frequency (time) scale is chosen to be the line-width (life-time) of the excited state $\gamma_{ei}$ (corresponding to $\gamma_{ei} = 5$ MHz). Accordingly, all frequencies can be converted to physical units. The classical fields are taken in the form of Gaussian pulses of peaks $\Omega_i$'s and widths $\sigma_{ci}$'s, overlapping each other in time. 

Transit of the atom through the overlapping waist region with a constant velocity $v_{at}$ leads to time-dependent cavity coupling coefficient $g_i$, taken as Gaussian pulses in simulations with widths $\sigma_{gi} \propto 1/v_{at}$ ( Fig.2 (a), (c), (e)). 

Figs. 2(a) and (b) correspond to splitting of a single photon equally in two cavity modes while Figs. 2(c) and (d) correspond to routing a photon from cavity $l$ to $r$. Both processes occur with unity efficiency. It is interesting to note that the classical fields are turned on before the transit of the atom, such that the two field shapes along with time-dependent cavity couplings overlap in time. This greatly simplifies the procedure without a need for fine tuning atomic velocity or pulse shapes.

Furthermore, one notes that the transfer occurs in a regime of $g_i$(= 3)$<<\Omega_{i} << \Delta$. However, for larger $g_i$ i.e. in the usual far-detuned limit of $g_i \geq \Omega_i \ll \Delta$ there is significant population accumulation in the intermediate states leading to partial transfer (Figs. 2(d) and (e)). Interestingly, this seems to indicate that a ``bad-cavity regime'' might be better suited for this model, as opposed to usual stringent ``strong-coupling'' requirement for single atom-cavity interactions. An analysis including decays (see below) supports this claim. Hereafter, we will refer to this regime as the \textit{beam-splitter regime} denoted by $g_i \ll \Omega \ll \Delta$. 

\begin{figure*}
\includegraphics[scale=0.63]{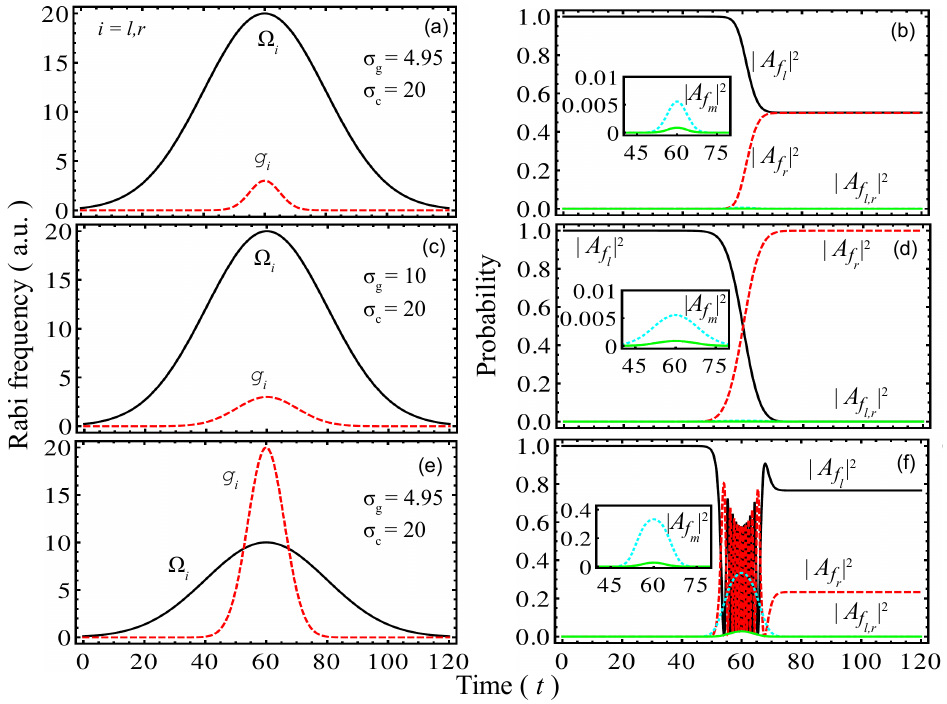}
\caption{Numerical simulation for the controlled beam-splitter: (a), (c), (e) depict Gaussian pulses of $\Omega_i$'s and   $g_i$'s in units of $\gamma_{ei} = 5$ MHz for three cases. The time-variations of $g_i$'s are due to the atom transiting with different velocities. (b), (d) and (f) are the corresponding variations of photon probabilities in two cavities $l$ and $r$. (b) corresponds to a half-transfer of population (with pulse sequences of (a)) from the dressed state $\ket{F_l}$, containing one photon in cavity $l$ to $\ket{F_r}$ with a photon in cavity $r$. The resulting photonic state is a coherent superposition of the two cavity modes. (d) depicts routing of a photon from cavity $l$ to cavity $r$ with near unity quantum efficiency. Insets of (b) and (d) show populations (two to three orders of magnitude smaller) of the intermediate ground state and excited states $\ket{e_i}$ and $\ket{f_m}$, respectively. In the regime of $g_i \geq \Omega_i$ corresponding to pulse shapes of (e), (f) shows poor routing efficiency, with significant occupation of the intermediate excited states, leading to spontaneous emission losses.}
\end{figure*} 

\textit{Beam-splitter regime:} 

The connection between a beam-splitter mixing two single mode optical fields with the Hamiltonian of Eq. (\ref{BS}) becomes evident if one compares the unitary for a beam-splitter with the effective unitary:
\begin{eqnarray} \label{Ueff}
\hat{U}_{eff} = \exp\left(-\frac{i}{\hbar}\int_0^{t_f}\hat{H}(t)dt \right)
= 
\begin{bmatrix}
A & B  \\
C & D  
\end{bmatrix},
\end{eqnarray} 
where $t_f$ denotes the duration of the pulse sequence. The elements $A$, $B$, $C$ and $D$ are time independent, though dependent strongly on the time dependence of $g_i(t)$ and $\Omega_i(t)$.
Numerical results show that by keeping the width and height of the classical pulses same and only changing the velocity of the transiting atom, one can continuously tune the elements of the unitary. 
Accordingly, 
we fix parameters ($g_i = 3, \Omega_i = 20$ and $\Delta_l=\Delta_r=\Delta = 50$) along with widths of the classical pulses ($\sigma_{ic} = 20$) and tune a single parameter, the transit velocity of the atom, thereby controlling all the elements of the unitary $\hat{U}_{eff}$. The corresponding numerical results for the matrix elements $A$ and $B$ as functions of velocity are presented in Fig. 3(a).

\begin{figure*}
\includegraphics[scale=0.6]{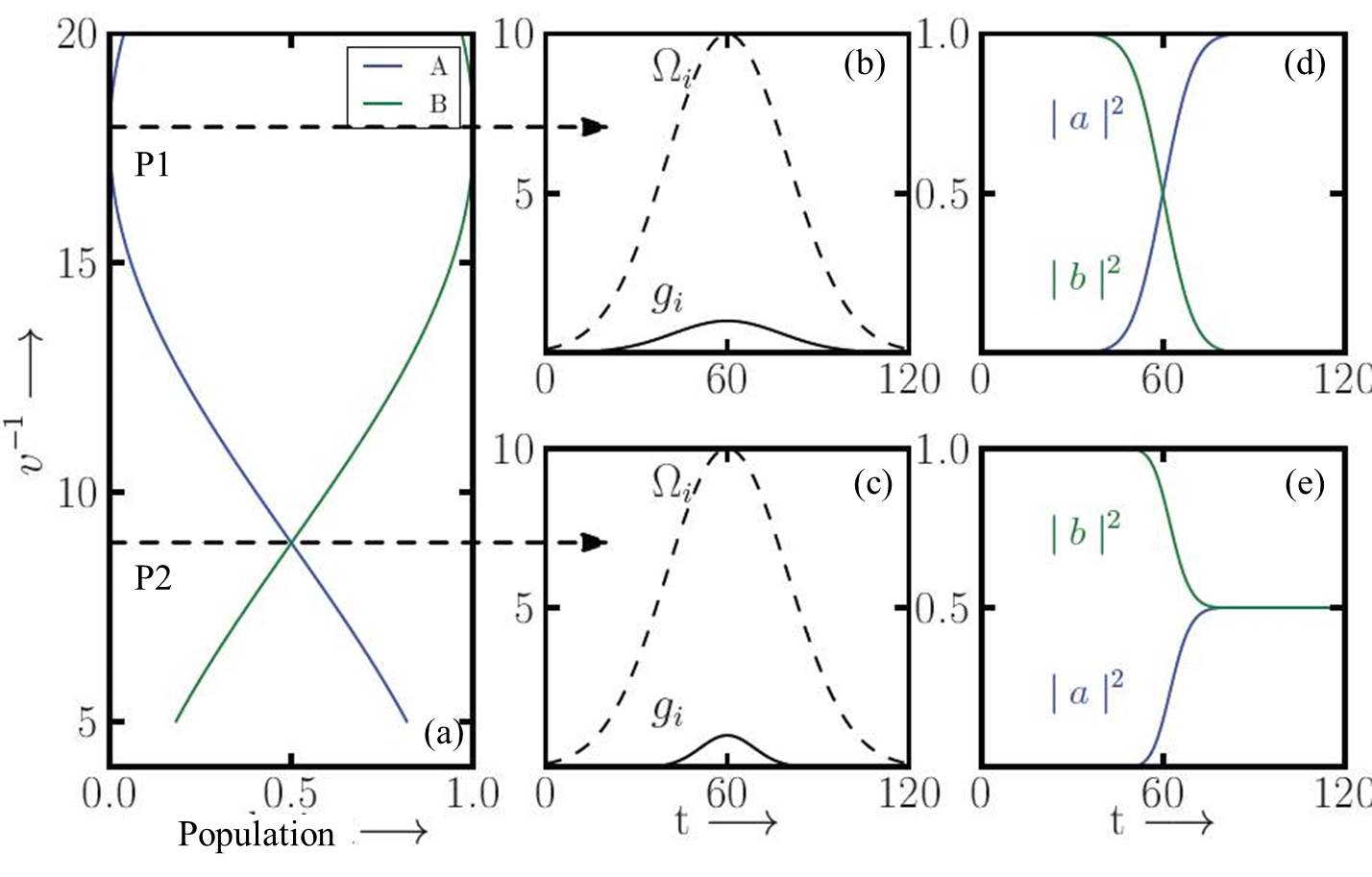}
\caption{Controlled Beam-splitter: (a) shows variations of matrix elements $A$ and $B$ of the unitary $\hat{U}_{eff}$ in Eq. (\ref{Ueff}), as the velocity of the transiting atom is varied. Two special points $P_1$ and $P_2$ are marked. $P_1$ corresponds to complete transfer of the photon from one cavity to the other. $P_2$ corresponds to equal-weighted superposition of two states of the photon. (b) and (d) show the variations of $\Omega_i$'s and  $g_i$'s corresponding to points $P_1$ and $P_2$ respectively. (c) and (e) show how the probabilities for the photon being in cavities $l$ and $r$ vary with time for $P_1$ and $P_2$ respectively.} 
\end{figure*} 

The point $P_1$ with $B=1.0$, corresponds to the photon-router regime with a single photon redirected from one cavity to the other. For cavities with a common and equal mode waist of $10 \mu m$ this corresponds to an atomic velocity of $20 m/s$ for the Rubidium atom. However, it is interesting to note that with a cavity waist of $100 \mu m$ one can get the same routing behaviour with a flying \textit{hot} atom with a velocity as high as $200 m/s$.
Realization of such a photon-router with hot atomic ovens without laser-cooling and trapping of single atoms can greatly simplify an important component of a future quantum network.

The point $P_2$ in Fig. 3(a) represents a superposition of a single photon being in either of the cavities, described as
\[
\ket{1_{l}, 0_{r}} \rightarrow \ket{1_{l}, 0_{r}} \pm \ket{0_{l}, 1_{r}}
\]
The usefulness of these photonic states as a $resource$ to perform quantum computation tasks comes due to their association with corresponding atomic states $(\ket{f_l,1,0}, \ket{f_r,0,1})$ as a \textit{marker}. Accordingly, a detection of a photon out of cavity $r$ together with a measurement of the atomic state uniquely characterizes the state $\ket{0_{l},1_{r}}$. 

We also note that the same set of operations can also be achieved when the atom is trapped in the waist, so that $g_i$'s do not vary. For example, an initial state $\ket{F_l}$ will adiabatically transform to $\ket{F_r}$ under the simultaneous variations of the form: $\Omega_r$ going from $\Omega_0 \rightarrow 0$ and $\Omega_l$ going from $0 \rightarrow \Omega_0$ with appropriate parameter variations (not shown here).

\section{Effect of Losses:} 

The single excitation can be lost either via a photon leaking out of the cavity or through spontaneous decay of the excited states in free-space modes. With the cavity decay rates as $\kappa_l$ and $\kappa_r$ and the decay rates of the excited levels $\ket{e_l}$ and $\ket{e_r}$ by $\gamma_l$ and $\gamma_r$ respectively, one can incorporate dissipation by modifying the diagonal elements of matrix $M$ (Eq.7) in an effective non-Hermitian Hamiltonian.

After adiabatically eliminating the excited states, cavity states $\ket{F_l}$ and $\ket{F_r}$ pick up an additional decay channel via the cavity couplings $g_i$'s. In the large detuning limit, since the excited states have amplitudes $\simeq g_i/\Delta_i$, the effective decay rates for the cavity states become $\Gamma_i = \kappa_i + \frac{\abs{g_i}^2}{\Delta_i^2}\gamma_i$, strongly suppressing spontaneous emission. Cavity states have one more decay channel through their coupling to the intermediate state $\ket{f_m,0,0}$ which has a linewidth. 
Though the intermediate state is a stable ground state, its acquires a width due to classical fields coupling it to  excited states, which decay. In the large-detuning limit, this coupling is $\approx \Omega_i /\Delta_i$, which gives a width $\sum_j \abs{\Omega_j}^2 \gamma_j / \Delta_i^2$ to the intermediate state. Recalling that the cavity states are coupled to the intermediate state by parameters $s_i$'s (definition following Eq. (9)) one obtains the total effective linewidths ${\Gamma_i}^{eff}, (i=l,r)$ as:
\begin{equation}
{\Gamma_i}^{eff} = \kappa_i + \frac{\abs{g_i}^2}{\Delta_i^2}\gamma_i + \frac{\abs{\Omega_i}^2\abs{g_i}^2}{\Delta_i^2 \sum_j\frac{\abs{\Omega_j^2}}{\Delta_j^2}\gamma_j}.
\end{equation}

The first term on the right, $\kappa_i$, is the \textit{desired} cavity decay channel. Furthermore, for the beam-splitter regime, $\kappa_i \ll \frac{g^2}{\Delta}$ (see the off-diagonal terms of Eq. (\ref{BS})) or equivalently, $\kappa_l,\kappa_r \ll g_l,g_r$. Physically, this condition requires the photon to be routed from one cavity to another before it decays out of the cavity. For the parameters used in Figs. 2 and 3, this requires $\kappa_i = 0.3$ which translates to cavities with coupling coefficient of $g_i \approx 5$ MHz and $\kappa_i \approx 500$ kHz.

The next two terms  are \textit{bad}, scattering the useful photon out in free-space (via the excited state). However, these scale as $\frac{\abs{g}^2}{\delta^2}$, implying  a scattering or failure rate of one in $10^4$ operations and can therefore be ignored.

Overall, the condition on the cavities, for the validity of the \textit{beam-splitter regime} including losses can now be stated as:
\[
\kappa_i \ll \gamma_i \approx g_i \\
\ll \Omega_i \ll \Delta .
\]
Accordingly, a reasonable set of parameters for Rubidium atoms and two identical cavities are:
$ \kappa = 500$ kHz, $g = 5$ MHz with classical field strengths $\Omega = 30$ MHz of pulse width 1 $\mu$s and a transiting atom of velocity $20$ m/s (for a cavity waist of 10 $\mu$m).

\section{A node of a quantum network}
We now show how this twin-cavity system can function as an elemental node in a quantum network. Firstly, one can note that the photon states for one node (with the two cavities) are also associated with atomic ground states $\ket{f_l}$ and $\ket{f_r}$. These two atomic states can therefore be ascribed to a qubit. The beam-splitter Hamiltonian, can then be used to produce any arbitrary superposition or qubit state. 

A quantum network can be envisaged to be a lattice of such nodes connected by optical fibers in an appropriate geometry. We consider two such nodes connected with a single-mode fiber. Specifically, arrangement considered here is one in which the $l$-cavity of one node is coupled to the $l$-cavity of the other as shown in Fig. 4. Next, we show how the two-node system described here can, in principle, perform, on demand, general one-qubit operations and some two-qubit gates in a controlled manner.

\begin{figure}[htbp]
\centering
{
\includegraphics[width=140mm]{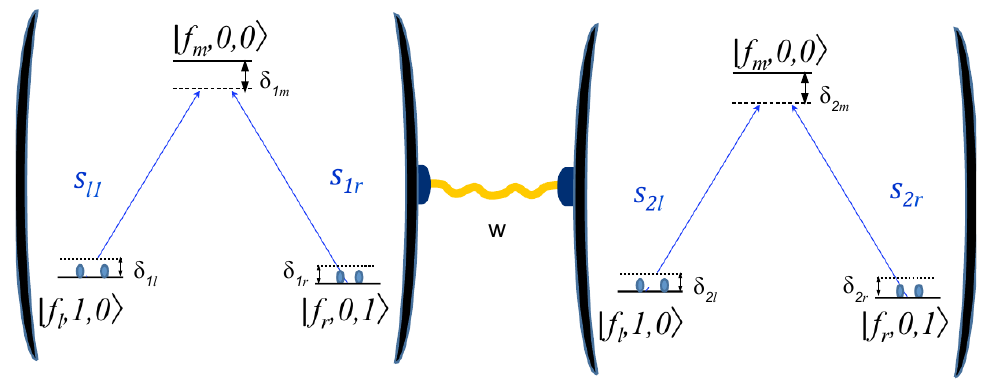}
}
\caption{Two atom-cavity system connected by a fiber. The level scheme shown is the truncated one, in which the excited levels of atoms are eliminated. It represents the Hamiltonian in Eq. (14).}
    \label{Fig2L}
\end{figure}

The Hamiltonian of this two-node system can be expressed as $H_2 =H_{node} + H_f$, where $H_{node}$ is the Hamiltonian of the two separate nodes and $H_f$ couples the nodes by transfer of photon through the fiber. 
With excitation number conserved, we once again limit the analysis to the single excitation sector. In
the large detuning-limit, the excited levels for each atom, $\ket{e_l}$ and $\ket{e_r}$, can be eliminated and we take the Stark shifts $\delta_i =\abs{g_i}^2/\Delta_i, i=l, r$ to be equal. The latter is not necessary but is algebraically simple for presentation.
One can then write,
\begin{eqnarray}\label{Hn}
H_{node} = \sum_{k=1}^2 [ \delta_k (\ket{F_l}_{k~k}\bra{F_l} + \ket{F_r}_{k~k}\bra{F_r} + \delta_{km} \ket{F_m}_{k~k}\bra{F_m}] \nonumber \\
+\sum_{k=1}^2 [s_{kl} \ket{F_l}_{k~k}\bra{F_m}  + s_{kr} \ket{F_r}_{k~k}\bra{F_m} + H.c.],
\end{eqnarray}
where $k$ labels the nodes and the subscript on states refer to the node number. Parameters for each node are written using obvious generalization of the earlier notation: $s_{ki}= g_{ki} \Omega_{ki}^*/ \Delta_{ki}$, $\delta_k = \abs{g_{ki}}^2/\Delta_{ki}$ for $k=1,2$ and $i=l,r$. Detuning $\delta_{km}$ is same as in Eq.(\ref{delta-m}) for node k. 

The fiber coupling, $H_{f}$ describes the transfer of the photon from the cavities to a mode in the fiber,
\begin{eqnarray}\label{Hf}
H_{f} = w \left[ b^{\dagger} (C_{1l} + C_{2l} + b (C_{1l}^{\dagger} + C_{2l}^{\dagger})\right],
\end{eqnarray}
where $b^{\dagger}$ creates a photon in the fiber mode, $w$ is the cavity-fiber transfer amplitude for the photon.

In this sector, the single photonic or atomic excitation can be in either of the nodes. So the possible states corresponding to single or zero excitations on one node is a larger set. We list these states with a shorter notation for convenience in Table I.

\begin{table}
\caption{One-node states making up the two-node wavefunction}
\begin{center}
    \begin{tabular}{ | p{3cm} |  p{3cm} |}
    \hline
    State Vector & Notation  \\ \hline
    $\ket{f_l, 0,0}$ & $\ket{l}$ \\ \hline
    $\ket{f_r, 0,0}$ & $\ket{r}$ \\ \hline
    $\ket{f_l, 1,0}$ & $\ket{F_l}$ \\ \hline
    $\ket{f_r, 0,1}$ & $\ket{F_r}$ \\ \hline
    $\ket{f_m, 0,0}$ & $\ket{F_m}$ \\ \hline
    $\ket{f_l, 0,1}$ & $\ket{u}$ \\ \hline
    $\ket{f_r, 1,0}$ & $\ket{v}$ \\ \hline
    \end{tabular}
\end{center}
\end{table}

In Table I, the first two states have zero excitation and the last 5 have one excitation. The Hilbert space for the two nodes consists of zero-excitation state of one node with one-excitation node of the other. In addition there are states in which the photon is in the fiber state. The two-node Hamiltonian connects $20$ such states which we label from 1 to 20 and write the two-node wave-function as
\begin{equation}
 \ket{\Psi_2} = \sum_{i=1}^{20} A_i(t) \ket{i}.
\end{equation}
The list of all such two-node states along with their notations is tabulated in Table II. The notation used is $\ket{a,b}\ket{}_f$ where a and b correspond to the states of node one and two respectively, and the second ket gives the photon number in the fiber. For writing the one-node states we use the shorter notation of Table I.

\begin{table}
\caption{Labels of Two-node states in single excitation sector}
\begin{center}
    \begin{tabular}{ | p{2cm} |  p{3cm} | p{2cm} |  p{3cm} |  }
    \hline
    Notation & State Vector  \\ \hline
    $\ket{1}$ & $\ket{F_l, l}\ket{0}_f$ & $\ket{2}$ & $\ket{F_l, r}\ket{0}_f$\\ \hline
    $\ket{3}$ & $\ket{F_r, l}\ket{0}_f$ & $\ket{4}$ & $\ket{F_r, r}\ket{0}_f$\\ \hline
    $\ket{5}$ & $\ket{F_m, l}\ket{0}_f$ & $\ket{6}$ & $\ket{F_m, r}\ket{0}_f$\\ \hline
    $\ket{7}$ & $\ket{l, F_l}\ket{0}_f$ & $\ket{8}$ & $\ket{r, F_l}\ket{0}_f$\\ \hline
    $\ket{9}$ & $\ket{l, F_r}\ket{0}_f$ & $\ket{10}$ & $\ket{r, F_r}\ket{0}_f$\\ \hline
    $\ket{11}$ & $\ket{l, F_m}\ket{0}_f$ & $\ket{12}$ & $\ket{r, F_m}\ket{0}_f$\\ \hline
    $\ket{13}$ & $\ket{l, l}\ket{1}_f$ & $\ket{14}$ & $\ket{l, r}\ket{1}_f$\\ \hline
    $\ket{15}$ & $\ket{r, l}\ket{1}_f$ & $\ket{16}$ & $\ket{r, r}\ket{1}_f$\\ \hline
    $\ket{17}$ & $\ket{l, v}\ket{0}_f$ & $\ket{18}$ & $\ket{v, l}\ket{0}_f$\\ \hline
    $\ket{19}$ & $\ket{r, v}\ket{0}_f$ & $\ket{20}$ & $\ket{v, r}\ket{0}_f$\\ \hline
    \end{tabular}
\end{center}
\end{table}

The Schr\"{o}dinger equation in terms of the coefficients can now be written straight forwardly using the Hamiltonian $H_2$ given in Eqs. (\ref{Hn}) and (\ref{Hf}).
\begin{equation}\label{Eqs}
 i \hbar \frac {d A_i}{d t} = \sum_j M_{ij} A_j.
\end{equation}
with the complete equations given in Appendix. It is observed that the large ($20X20$) matrix $M_{ij}$ is rather sparse. A closer inspection reveals that the states are divided into groups of states which are just connected within each group. For example, states $\ket{2}, \ket{4}, \ket{6}, \ket{14}, \ket{17}$ form such a group. Certain eigenstates of these groups have characteristics of dark states, which allow adiabatic rotations among states of interest \cite{Pellizzari97,Kumar12}. The case $\delta_1 = \delta_2 = 0$ algebraically has the simplest structure and one can identify three degenerate eigenstates of zero energy for qubit operations. These are:

\begin{equation}
 \ket{D_3} = \left[\ket{F_l, r} - \frac{s_{1l}}{s_{1r}} \ket{F_r, r} - \ket{l, v} \right]\ket{0}_f,
\end{equation}

\begin{equation}
 \ket{D_4} = \left[\ket{r, F_l} - \frac{s_{2l}}{s_{2r}} \ket{r, F_r} - \ket{v, l}\right]\ket{0}_f,
\end{equation}

\begin{equation}
\ket{D_5} =\left[\frac{1}{s_{1r}}(s_{1r}\ket{F_l, l} - s_{1l}\ket{F_r, l} ) - \frac{1}{s_{2r}}(s_{2r}\ket{l, F_l} - s_{2l}\ket{l, F_r}) \right] \ket{0}_f.
\end{equation}

These states have several interesting feature. For example, for $\ket{D_5}$, adiabatic rotations in only one node (say vary $s_{1l}$ and $s_{1r}$) results in a intra-node transfer of a photon from one cavity to the other, while the state of the other node remains completely unaffected. For general detuning, these states are more complex and such adiabatic-rotation analysis, serves only as a broad guide. One can accordingly, resort to numerical simulations.

We have done numerical simulations for the complete set of 20 equations Eq. (\ref{Eqs}). Various protocols for parameter variations were tried and we present some in which useful operations can be done with high fidelity. It can be noted that there are several free parameters in the system. For two nodes we have four laser pulses of Gaussian shape whose amplitudes ($\Omega_0$) and widths ($\sigma_c$) can be varied. Similarly there are four atom-cavity couplings, which are again taken to be Gaussian with respective amplitudes $g_0$ and widths $\sigma_g$. The time interval between the pulses can also be varied. The frequency (time) scale is the same as in the last section, i.e. spontaneous decay rate of the excited atomic levels. As far as possible, parameters are kept the same in all the simulations. These are: $\Delta_{1l} = \Delta_{2l} = \Delta_{1r} = \Delta_{2r} = \Delta$ = 50. while typical laser pulse magnitudes are  $\Omega_{0}$ = 20, atom-cavity coupling $g_0$ = 3 and pulse widths: $\sigma_c$ = 20, $\sigma_{g}$ = 10. Any finer changes around these values are mentioned in figures. 

\subsection{One-qubit operations} Single qubit operations for an isolated node have already been demonstrated in the earlier sections. Here, we show that even for a two-node system with weak fiber-coupling, single qubit operations can be performed on one node without disturbing the other qubit. Figure 5(a) shows a set of laser pulses ($\Omega_{1l}$ and $\Omega_{1r}$) applied on node-1, along with a weak variation of coupling parameters $g_{1i}$. Figure 5(b) shows that only two amplitudes change, resulting in a transfer of excitation from the initial state $\ket{F_r, r}$ to $\ket{F_l, r}$, while all other state amplitudes at both nodes remaining close to zero. Furthermore, small variations in parameters for pulses on node-1 (Fig. 5(c)) results in another important one-qubit operation (Fig. 5(d)). Here one generates an equal superposition $\ket{F_r, r} + \ket{F_l, r}$ from an initial state $\ket{F_r, r}$. In the lower most panel we show a flip operation on node-2. Figures 5(e, f) show the pulse and the corresponding variations of amplitudes in which the initial state $\ket{r, F_r}$ is transformed to $\ket{r, F_l}$. Note with a low fiber coupling $w=0.001$, all these operations have high fidelity $\approx 1$ (in absence of dissipation). Though with increased coupling the fidelity drops, nevertheless a large parameter space remains to be explored for optimal operations.

\begin{figure*}
\includegraphics[scale=0.6]{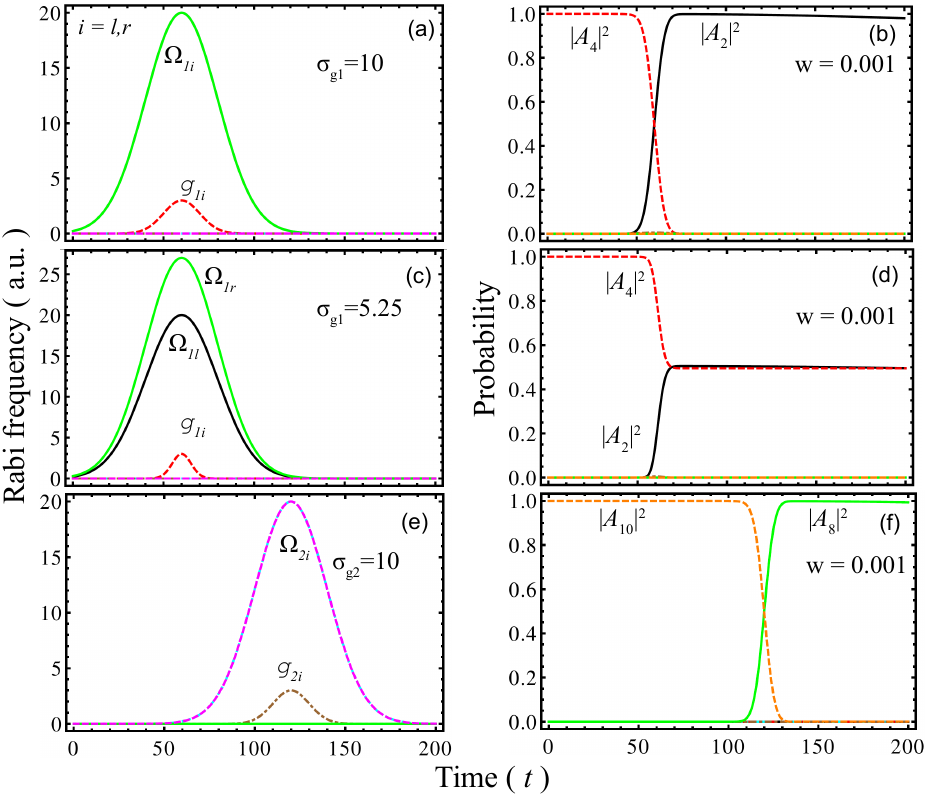}
\caption{One-qubit operations for the two-node system. (a) shows a pulse protocol on node-1 which flips the qubit on the node. This is shown in (b) where the state $\ket{F_r,r}$ is transformed to $\ket{F_l,r}$. (c) shows a pulse protocol on node-1  which as seen in (d) generates a superposition $\ket{F_r,r}+ \ket{F_l,r} $ on the node from an initial state $\ket{F_r,r}$. Note parameter changes here: $\Omega_{0l}$ = 20, $\Omega_{0r}$ = 27 and $\sigma_{g1}$ = 5.25. (e) shows a pulse-protocol on node-2 which results in flipping qubit on node-2 as shown in (f). Here the initial state $\ket{r, F_r}$ is transformed to $\ket{r, F_l}$. }
\end{figure*} 

\subsection{Two-qubit operations}
Next we demonstrate few two-qubit operations. As seen above, all qubit operations require presence of a photon in a node. Since we are in the single-photon sector, it is essential to be able to transfer a photon from one node to the other. The top panel of Fig. 6 shows such an operation. Fig. 6(a) shows two sets of pulses applied on node-1 and node-2 successively. Fig. 6(b) shows the variation of amplitudes for the resulting process. Here the initial state $\ket{F_r, l}$ is transformed to $\ket{l, F_r}$, but it should be noted that the amplitude for the state $\ket{l, F_l} + \ket{F_l, l} $ becomes substantial after the first set of pulses applied on node-1. This is to be expected as the photon transfers to the fiber from $l$-cavity of node-1 and then to the $l$-cavity of node-2. The photon transfer in conjunction with single node opearations can yield many useful gates required in quantum processing. In the middle panel, Fig. 6(c) shows a sequence of two pulses applied successively on nodes 1 and 2. The corresponding variations of amplitudes in 6(d) show that the initial state $\ket{F_r, l}$ is transformed to $\sqrt{2}\ket{l, F_r} + \ket{r, F_l}$ via the intermediate state $\ket{l, F_l} + \ket{F_l, l} $. This is an example of an entangled state of two qubits. Figs. 6(e) and (f) show another operation in which the first pulse on node-1 transforms the state $\ket{F_r, l}$ to a superposition $\ket{l, F_l} + \ket{F_l, l} + \ket{l, F_r}$. The second pulse on node-2 then transforms this state to $\sqrt{2}(\ket{F_r, l} + \ket{l, F_r}) + (\ket{l, F_l} + \ket{F_l, l}) $. In terms of atomic qubits this state is $\ket{r, l} + \ket{l, r} + \ket{l, l} $. Note that in these operations, common parameters are: $\Omega_{0l}$ = 20, $\Omega_{0r}$ = 27, $\sigma_{c2}$ = 20, $w$ = 0.6 and time interval between the pulses is 150. Other variations are mentioned in figures.

\begin{figure*}
\includegraphics[scale=0.6]{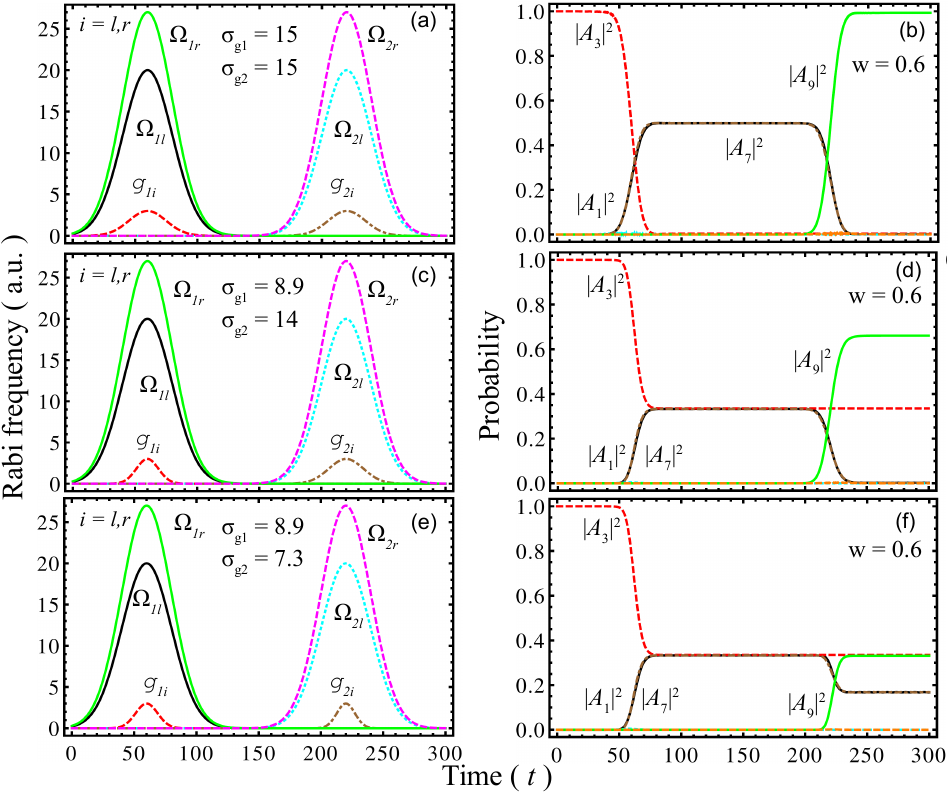}
\caption{Three operations involving transfer of a photon from one node to the other. In the top panel (a) shows a sequence of two pulses applied on node-1 and node-2 that results (b) in transfer of the state $\ket{F_r, l}$ to $\ket{l, F_r}$. In the middle panel (c) shows a small variation in widths of the two pulses that results (d) in transfer of the state $\ket{F_r, l} $ to $\sqrt{2}\ket{l, F_r} + \ket{r, F_l}$. This is an entangled state of two qubits. The bottom panel (e) and (f) show a procedure in which the first pulse on node-1 transforms the state $\ket{F_r, l}$ to a superposition $\ket{l, F_l} + \ket{F_l, l} + \ket{l, F_r}$. The second pulse on node-2 then transforms this state to $\sqrt{2}(\ket{F_r, l} + \ket{l, F_r}) + (\ket{l, F_l} + \ket{F_l, l}) $. }
\end{figure*}

Fig. 7 shows two protocols with the initial state $\ket{l, F_l} + \ket{F_l, l}$. In the upper panel Figs. 7(a) and 7(b) a set of two pulses applied on node-1 and node-2 successively leads to its transformation to $\ket{F_r, l}$. While in the lower panel both the qubits are flipped resulting in the state $\ket{l, F_r} + \ket{F_r, l}$ by a variation in the parameters of the two pulses. The common parameters used here are same as in Fig. 6.
The operations shown here are some representative examples which demonstrate the flexibility and versatility of the scheme.

\begin{figure*}
\includegraphics[scale=0.6]{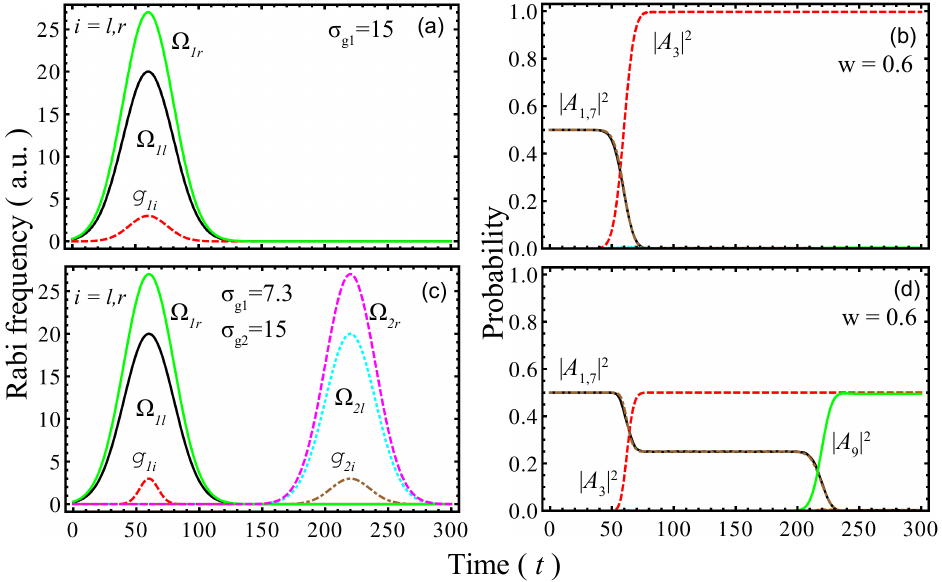}
\caption{Two operations on a initial entangled state $\ket{l, F_l} + \ket{F_l, l}$ of two nodes are shown. In the upper panel a single pulse on node-1 transforms the state to $\ket{F_r, l}$. In the lower panel a simultaneous flip on both the qubits occurs resulting in the state $\ket{l, F_r} + \ket{F_r, l}$. Parameters used are same as common parameters mentioned for Fig. 6. The changes in pulse widths are depicted.} 
\end{figure*} 

\section{Conclusion}

To conclude, we have proposed a model for a node of a quantum network consisting of two electromagnetic cavities with overlapping waists. The cavities are coupled to an effective five-level atom. The node performs a number of quantum operations in a controlled manner. Our first application of the node is demonstrated through routing photons with unity efficiency and as a controllable beam-splitter. We demonstrate these two operations when the coupling is provided by flying atoms through the waist. The key advantage of the scheme is that it does not place stringent requirements on pulse sequences or atom-cavity coupling, and is workable with reasonably achievable cavity parameters and atom velocities.

As applications relevant for a future quantum network, we have considered a system of two nodes coupled by an optical fiber and shown how some elemental one and two qubit operations can be performed. Qubits for these operations are degenerate atomic states while operations are performed by photon transfer within two cavities of a node or via inter-node transfer. Explicit parameter regimes on laser powers and atom cavity couplings are presented for (a) unitary qubit operations on one node without disturbing the state of the other qubit, (b) generating a two node superposition and effecting a swap operation, (c) generating entangled Bell states of two qubits.

The proposed model has been analysed in a cavity QED scenario. However, the protocols can be applied to several other physical systems. In particular, coupled photonic crystal cavities along with gate-induced tuning of a single coupled quantum dot\cite{AK13} or tunable superconducting qubits coupled to multiple microwave resonators\cite{Sch07} can also lead to viable possibilities of experimental implementation of the proposed node in future.

\section*{Acknowledgements} SG acknowledges support from DST-SERB(SB/S2/LOP-05/2013) and DK is grateful for the support of Raja Ramanna fellowship of the Department of Atomic Energy, Government of India.

\section*{Appendix}
Here we write the full form of 20 equations of motion Eq. (17) for the coupled two-node system. They are obtained directly using the Hamiltonian given in Eq. (14). The notation for the states is provided in Tables I and II. Since many equations are very similar, we have combined two or three together.
\begin{eqnarray}
i \dot{A}_{1,2} &=& \delta_1 A_{1,2} + s_{1l} A_{5,6} + w A_{13,14} \nonumber \\
i \dot{A}_{3,4} &=& \delta_1 A_{3,4} + s_{1r} A_{5,6} \nonumber  \\
i \dot{A}_{5,6} &=& s_{1l}^* A_{1,2} + s_{1r}^* A_{3,4} + \delta_{1m} A_{5,6} \nonumber  \\
i \dot{A}_{7,8} &=& \delta_{2} A_{7,8} + s_{2l} A_{11,12} + w A_{13,15} \nonumber \\
i \dot{A}_{9,10} &=& \delta_{2} A_{9,10} + s_{2r} A_{11,12} \nonumber \\
i \dot{A}_{11,12} &=& s_{2l}^* A_{7,8} + s_{2r}^* A_{9,10} + \delta_{2m} A_{11,12} \nonumber \\
i \dot{A}_{13} &=& w (A_{1} + A_{7}) \nonumber \\
i \dot{A}_{14,15,16} &=& w (A_{2,8,19} + A_{17,18,20}) \nonumber \\
i \dot{A}_{17,18} &=& w A_{14,15} \nonumber \\
i \dot{A}_{19,20} &=& w A_{16} \nonumber
\end{eqnarray}
One can immediately see that $A_{16}, A_{19}$ and $A_{20}$ are just coupled amongst themselves and involve states of no physical interest here. The results in the text are obtained by solving the remaining set of 17 equations numerically with time variations of the coupling coefficients as indicated in the text.

\end{document}